\documentclass[aps,prb,twocolumn,groupedaddress,amssymb,superscriptaddress]{revtex4}
\usepackage[dvips]{graphicx}
\DeclareGraphicsExtensions{.eps,.ps}
\usepackage{bm}      
\newcommand{\lfaof}{La\-Fe\-As\-O$_{1-x}$\-F$_x$}
\newcommand{\sfaof}{Sm\-Fe\-As\-O$_{1-x}$\-F$_x$}

\newcommand{\erf}{\mbox{\rm erf}}
\begin{document}
\bibliographystyle{apsrev}

\title{Magnetic-superconducting phase boundary of SmFeAsO$_{1-x}$F$_{x}$ \\
           studied via muon spin rotation: Unified behavior in a pnictide family}

\author{S.~Sanna}
\email[]{Samuele.Sanna@unipv.it}
\affiliation{Unit\`a CNISM di Parma e Dipartimento di Fisica, I 43100 Parma, Italy}
\affiliation{Dipartimento di Fisica "A.Volta" e Unit\`a CNISM di Pavia, I 27100 Pavia, Italy}
\author{R.~De Renzi}
\affiliation{Unit\`a CNISM di Parma e Dipartimento di Fisica, I 43100 Parma, Italy}
\author{G.~Lamura}
\affiliation{CNR-INFM-LAMIA and Universit\`a di Genova, via Dodecaneso 33, 16146 Genova, Italy}
\affiliation{CNR-INFM Coherentia, Università di Napoli Federico II, 80125, Napoli, Italy.}

\author{C.~Ferdeghini}
\affiliation{CNR-INFM-LAMIA and Universit\`a di Genova, via Dodecaneso 33, 16146 Genova, Italy}
\author{A.~Palenzona}
\affiliation{CNR-INFM-LAMIA and Universit\`a di Genova, via Dodecaneso 33, 16146 Genova, Italy}
\author{M.~Putti}
\affiliation{CNR-INFM-LAMIA and Universit\`a di Genova, via Dodecaneso 33, 16146 Genova, Italy}
\author{M.~Tropeano}
\affiliation{CNR-INFM-LAMIA and Universit\`a di Genova, via Dodecaneso 33, 16146 Genova, Italy}
\author{T.~Shiroka}
\affiliation{Laboratory for Muon-Spin Spectroscopy,
Paul Scherrer Institut, CH-5232 Villigen PSI, Switzerland}

\date{\today}

\begin{abstract}
We present $\mu$SR investigations on \sfaof\ showing coexistence of magnetic order and superconductivity only in a very narrow F-doping range. The sharp crossover between the two types of order is similar to that observed in \lfaof, suggesting a common behavior for the 1111 pnictides. The analysis of  the muon asymmetry demonstrates that the coexistence must be nanoscopic, i.e.\ the two phases must be finely interspersed over a typical length-scale of few nm.  In this regime both the magnetic and the superconducting transition temperatures collapse to very low values. Our data suggest a competition between the two order parameters.

\end{abstract}


\maketitle




The interplay of superconductivity and magnetism in the recently discovered Fe pnictides\cite{Kamikara} is a controversial issue. Most of the parent compounds of Fe-pnictides are magnetically ordered. Upon doping, magnetism disappears and superconductivity emerges, but the nature of this crossover remains unclear. A direct competition between the two order parameters, leading to the coexistence of interspersed nanoscopic domains, is the hallmark of unconventional superconductivity. It has been observed in cuprates,\cite{Savici,Sanna04,Russo} and directly imaged by STM.\cite{Kohsaka} Also ruthenocuprates\cite{Tokunaga}, e-doped cuprates\cite{Kang} and heavy fermion superconductors\cite{Llobet,Park,Visser} manifest similar characteristics on a microscopic scale.

Establishing a similar occurrence for Fe pnictides would strongly indicate their departure from conventional superconductivity. However, the experimental evidence is controversial and the answer seems to vary across the different pnictide  families.
Muon spectroscopy ($\mu$SR) is often employed in these studies, thanks to its spectroscopic ability to combine the identification of different local surroundings of the implanted muons (i.e.~the specific signature) with the quantitative determination of the sample volume fraction giving rise to that signature. At one extreme, in the \lfaof\ system, a first-order-like transition between the magnetic (M) and the superconducting (SC) region of the phase diagram is observed by $\mu$SR,\cite{Luetkens} with no coexistence and very little doping dependence on both transition temperatures, $T_m$ and $T_c$, respectively.
At the other end a marked dependence of $T_m$ and $T_c$ on doping is reported in \sfaof\, \cite{Drew,Margadonna,Liu} as well as in AFe$_2$As$_2$ compounds (A=Ba,Sr,Ca).\cite{Goko,Laplace}  \sfaof\ has been reported to display peculiar features, with an unusually high crossover concentration and a wide interval, $0.1\!\!\le\!\! x\!\!\le$0.15, where superconductivity coexists with magnetism.\cite{Drew} This coexistence is shown to occur at a nanoscopic level, in contrast to other pnictides.\cite{Goko,Julien} Surprisingly, sizable volume fractions show a fluctuating rather than a static magnetic state.\cite{Drew} A possible influence of sample inhomogeneity is suggested explicitly in the accompanying comment.\cite{Uemura} Neutron data on CeFeAsO$_{1-x}$F$_x$ evidence a smooth variation of the two order parameters, which are both suppressed at the crossover concentration, suggesting the existence of a quantum critical point,\cite{Zhao} which, however, is not confirmed by $\mu$SR.\cite{Amato} One must be also aware that for systems containing magnetic rare earths further phenomena take place at very low temperatures (rare earth moment fluctuations and ordering).\cite{Khasanov} A more detailed comparison among different families may be found in a recent review,\cite{Amato} which also concludes by invoking further investigation.

We present a set of data on \sfaof (Sm-1111), on samples with rather sharp magnetic and superconducting transitions. Although they do support a well defined instance of nanoscopic coexistence, they display a rather abrupt crossover between the M and the SC phases, confined to an extremely narrow doping range, virtually point-like,  which is compatible with that observed in La-1111 compounds.\cite{Luetkens} Our findings suggest the existence of a common behavior at the crossover for all 1111 oxypnictides, not dependent upon the rare earth ion, in contrast with earlier views.\cite{Drew,Zhao}

Our polycrystalline Sm-1111 samples were synthesized in sealed crucibles of tantalum
\cite{Tropeano}. This procedure reduces F losses, since it avoids the partial reaction
of fluorine with the quartz vessel. It guarantees that the doping content strictly scales
with the nominal one, $x$, which is intended both as an upper limit to the real content and as a sample label.
In this paper we focus on samples close to the M-SC crossover. We investigate their superconducting behavior by field-cooling (FC) SQUID and transverse field (TF) $\mu$SR, while the presence of a magnetically ordered phase is detected by zero-field (ZF) $\mu$SR.

\begin{figure}
\includegraphics[width=0.38\textwidth,angle=0]{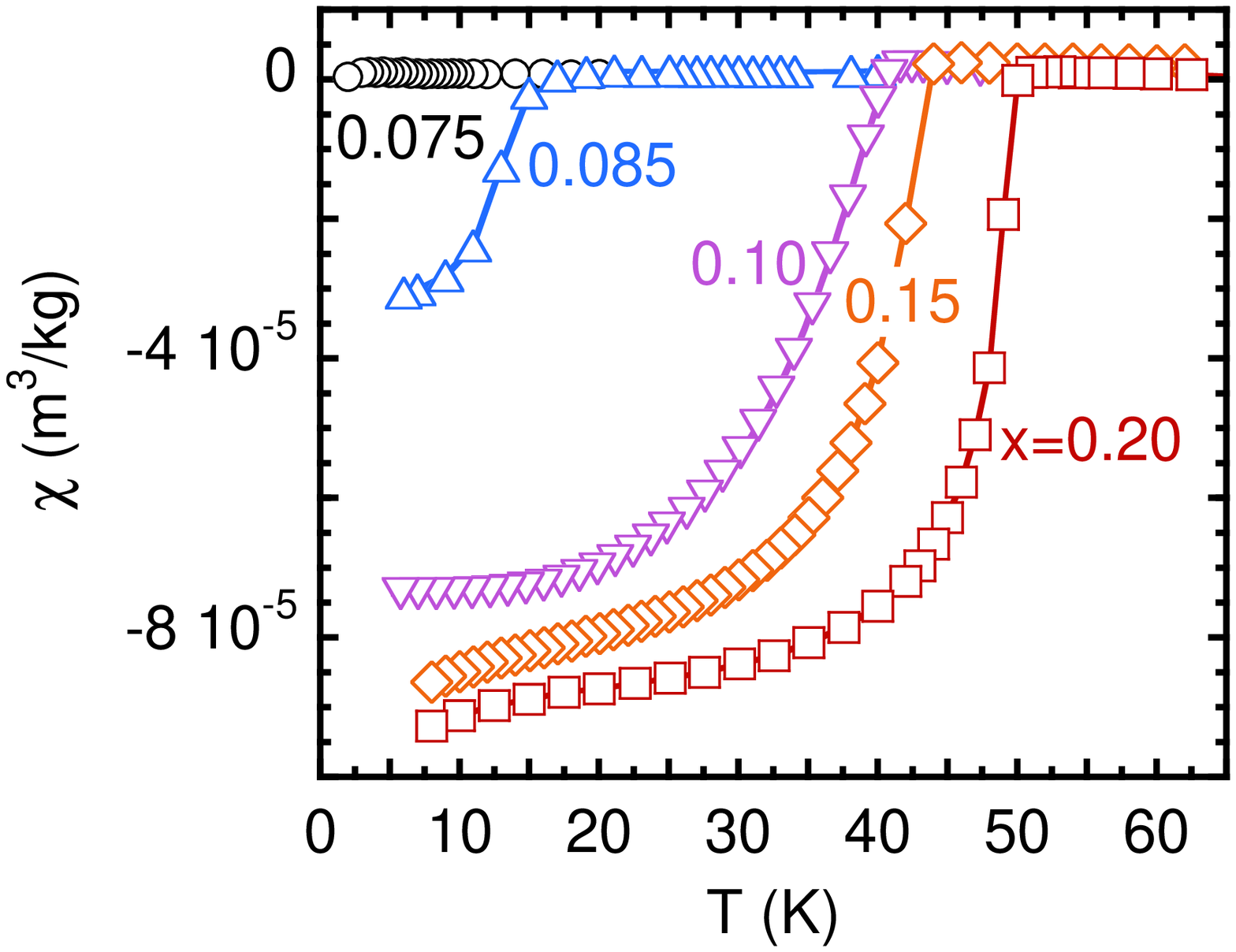}%
 \caption{ (Color online) Susceptibility of \sfaof\ from SQUID measurements in $\mu_0H=0.3$ mT (field-cooling).}
 \label{fig:squid}
\end{figure}

Figure~\ref{fig:squid} shows the sample susceptibility as a function of temperature, as measured by SQUID magnetometry in low-field FC experiments. A small contribution from marginal ferromagnetic (FM) inclusions with Curie temperatures in excess of 300 K was subtracted as a constant background.
These FC results provide a lower limit for the superconducting volume fraction. Assuming bulk superconductivity at optimal doping, $x$=0.20, in first approximation the SC fraction is given by the ratio
$v(x)=\chi(x)/\chi(0.20)$ at low temperature. The ratio remains nearly constant
down to $x$=0.10, and it is still large ($v=0.3$) for the sample $x$=0.085, where we expect, however, that the value is underestimated. Indeed, at low doping there is a
considerable increase\cite{Carlo} of the field penetration depth $\lambda_L$, which
becomes comparable to the grain dimensions (1-10 $\mu$m). The resulting reduction in
the Meissner volume of each grain is then reflected in a lower $\chi(x)$
value and hence in an underestimated SC volume fraction.

For the $x$=0.085 sample the transition ($T_c$=14(1) K, extrapolated linearly from the 10-90\% diamagnetic drop of $\chi$, which occurs within $\Delta T=5$ K) is rather narrow, much narrower than in samples with comparable $T_c$ reported elsewhere. Since the $x$=0.075 sample shows no bulk superconductivity (Fig.~\ref{fig:squid}), we conclude that the M-SC crossover occurs close to $x$=0.08, at a much lower doping than in previous reports.

Assessing the bulk character of the SC phase is a typical task for TF-$\mu$SR. When the sample is field-cooled below $T_c$ in a magnetic field $\mu_0\bm{H}$ orthogonal to $\bm{S}_\mu$, the Abrikosov flux lattice (FL) is formed.  The FL produces a distribution of internal fields $B(\bm{r})$ that is uniformly probed by the precessing muons, whose lineshape provides the moments of the distribution.\cite{Brandt}

\begin{figure}
\includegraphics[width=0.36\textwidth,angle=0]{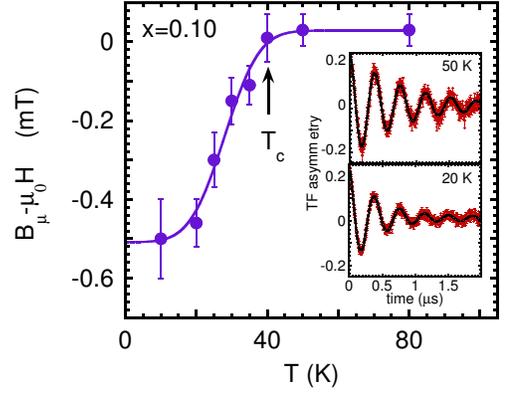}%
 \caption{(Color online) Internal field $B_\mu(T)-\mu_0 H$ from TF $\mu$SR ($\mu_0 H\!\!=\!\!20$ mT) for \sfaof\ at $x$=0.10. Insets: time dependence of asymmetry above and below $T_c$; solid curves are best fits (see text). }
 \label{fig:TF}
\end{figure}

\begin{figure}
\includegraphics[width=0.47\textwidth,angle=0]{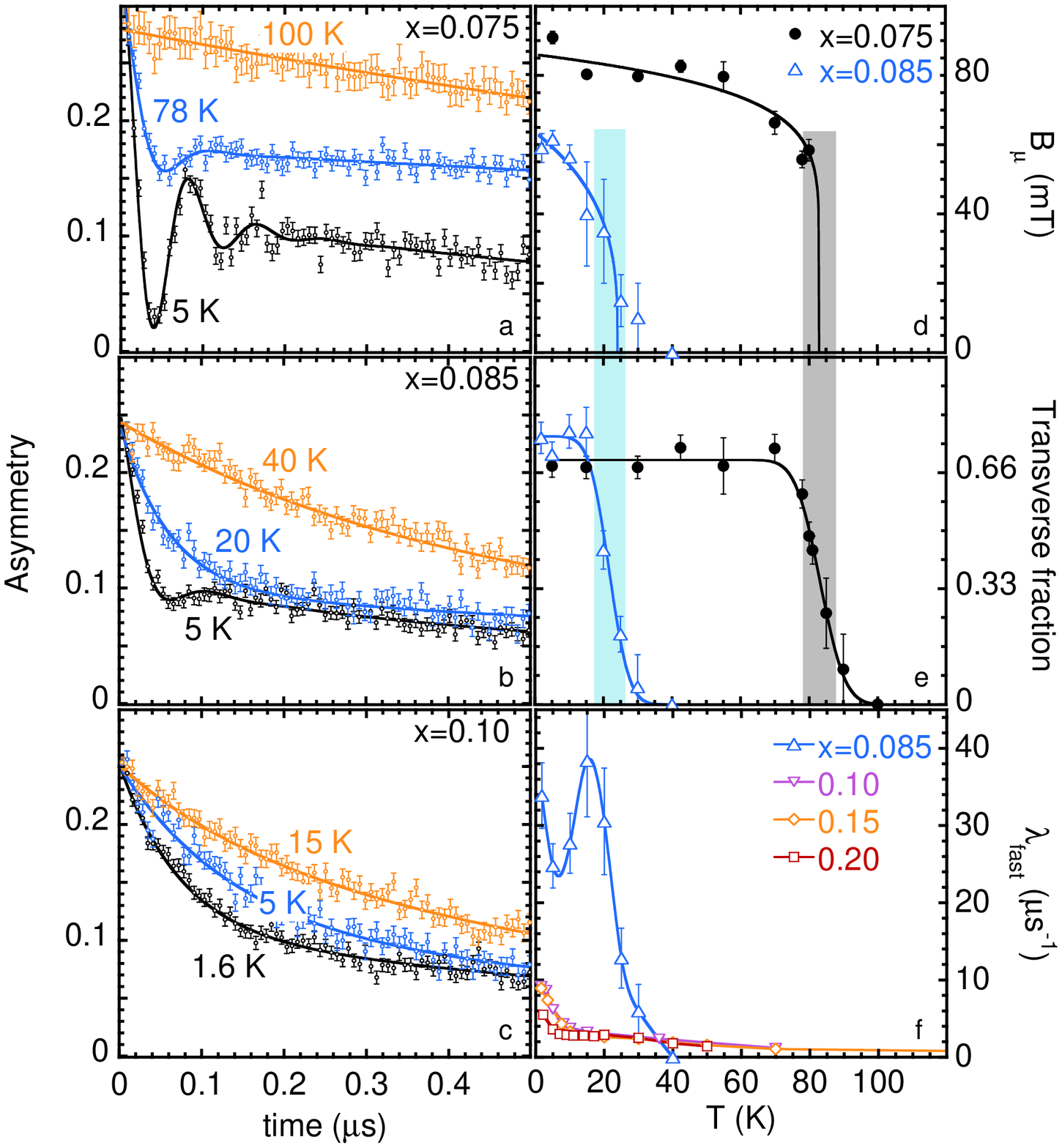}%
 \caption{(Color online) ZF $\mu$SR of \sfaof. Representative muon asymmetries for a) $x$=0.075, b) $x$=0.085 and c) $x$=0.10.  d) Internal field and e) transverse fraction $a_T/a_{ZF}$ with their best fit (Eq.~\ref{eq:ft}); f) relaxation rates, $\lambda_\mathrm{fast}$ for the fully SC samples and $\lambda_T$ for $x$=0.085, as a function of temperature (solid lines are guides to the eye).}
 \label{fig:ZF}
 \end{figure}

This type of measurement is not possible in the $x$=0.085 sample, where the ZF-$\mu$SR  response is dominated by the large internal magnetic fields which set in below 22 K, i.e.\ well before the SC transition ({\em vide infra}, Fig.~\ref{fig:ZF}d).
The detection of the FL is however possible for $x\!\!\ge$0.10. The time dependence of the TF asymmetry in the least doped of these samples, $x$=0.10, is shown in the inset of Fig.~\ref{fig:TF} for two temperatures, above and below the SC transition as detected by SQUID (onset $T_c=41(1)$ K). The  best fit of the asymmetry data to ${\cal A}_T (t) \!\!=\!\! a_{TF}  \exp(-\lambda_T t)\cos(2\pi\gamma B_\mu t) $ is also shown, where $a_{TF}$ is the amplitude and $\lambda_T$ the relaxation rate. Figure~\ref{fig:TF} shows the shift of the internal field $B_\mu(T)-\mu_0 H$ as a function of  temperature, displaying a clear diamagnetic behavior below $T_c$, in agreement with the SQUID data. This indicates that a large fraction of muons detect a field lower than the applied one, due to the screening of the applied field in the bulk superconductor.

The  TF-$\mu$SR results confirm the bulk character of superconductivity for $x\ge$0.10. They support the assumption ($v\approx 1$ for $x$=0.2) that we initially made to estimate the superconducting volume fractions $v(x)$ from SQUID susceptibility.

We consider now magnetic order, detected by means of ZF-$\mu$SR. The left panels in Fig.~\ref{fig:ZF} display  the time dependence of the ZF muon asymmetry, ${\cal A}_Z(t)$, for the three samples close to the M-SC phase boundary, $x$=0.075, 0.085 and 0.10 at three different temperatures each. The solid lines show the best fit of the sample asymmetry  to the following standard ZF function:
\begin{equation}
{\cal A}_Z(t) = a_T \cos(2\pi\gamma |{\bm{B}_i}| t) e^{-\lambda_T t} + a_L G(t),
\label{eq:Zasymmetry}
\end{equation}
where $\gamma\!\!=\!\!135.5$ MHz/T is the muon gyromagnetic ratio,  $\bm{B}_i$ the magnetic field at the muon site, $\lambda_T$ a relaxation rate and $G(t)$ a non-oscillating relaxation function. At the highest temperature shown all three samples are in the paramagnetic regime and $a_T$ vanishes together with $B_i$. By converse, at lower temperatures both the $x$=0.075 and the $x$=0.085 samples display damped oscillations, due to the presence of internal fields of the order of 100 mT, compatible with a spin density wave (SDW) ordering of the FeAs layers.\cite{Drew,Goko} The initial direction of the muon spin $\bm{S}_\mu$ distinguishes the transverse precessing component ($a_T$, $\bm{B}_i\perp \bm{S}_\mu$) and the longitudinal component ($a_L$, $\bm{B}_i\!\!\parallel\!\! \bm{S}_\mu$), whose powder-average amplitudes yield a constant sum $a_T+a_L=a_{ZF}$.
Owing to the large damping, fits with Bessel functions (the expected SDW lineshape) yield $\chi^2$ values comparable  to those with a cosine fit, precluding the distinction between SDW and simple antiferromagnetic (AF) order.
The longitudinal component reveals two contributions, distinguished by largely different relaxation rates, $a_LG(t) = a_1 e^{-\lambda_1} + a_2 e^{-\lambda_2}$, up to almost room temperature. They are due to the already mentioned distant diluted ferromagnetic inclusions, observed also by SEM analysis\cite{Martinelli} and NMR,\cite{Sidorenko} with a total volume fraction too small to be detected directly.

Figures \ref{fig:ZF}d and \ref{fig:ZF}e show the internal field, $B_i$, and the transverse fraction, $a_T/a_{ZF}$, as a function of temperature. The simultaneous vanishing of the internal fields $B_i$ and of the transverse amplitudes $a_T$ identify the magnetic transition temperature,
$T_m$, which is obtained from a power-law fit, $B_i(T)=B_{0}(1-T/T_m)^\beta$.
If all the muons experience a net internal field $\bm{B}_i$, a simple geometric argument predicts  $a_T/a_{ZF}$=2/3 for polycrystalline samples and a more accurate determination of $T_m$ is obtained by fitting the transverse component to:
\begin{equation}
\label{eq:ft}
a_T(T)={a_T(0)\over {2 a_{ZF}}}\left[1-\erf\left({{T-T_m}\over{\sqrt2\Delta T_m}}\right)\right],
\end{equation}
which reproduces a Gaussian distribution of local transition temperatures around the average value $T_m$. The best fit yields $T_m$=83 K and 22 K for $x$=0.075 and 0.085 respectively, both with a RMS deviation $\Delta T_m$=5 K, indicating relatively sharp transitions.

If extensive parts of our $x\le 0.085$ samples were not magnetically ordered their transverse component would be significantly less than 2/3. This component would also be reduced in the presence of clusters with fluctuating magnetization, such as have been reported in earlier Sm-1111 work\cite{Drew} (their origin was possibly extrinsic, since we definitely could not detect them). Figure~\ref{fig:ZF}e shows that below $T_m$ our measured transverse component is close to the ideal value of 2/3, demonstrating the presence of static moments throughout the {\em whole} volume for both the $x=0.075$ and 0.085 samples. These must be Fe static moments since the observed field values of $\sim100$ mT is the signature of the ordered Fe sublattice.\cite{Luetkens}

The sharp magnetic transitions for $x\le 0.085$ are also reflected in the critical peak of the transverse relaxation rate $\lambda_T$, as shown in Fig.~\ref{fig:ZF}f in the case of $x$=0.085. This peak is due to the slowing down of the spin fluctuations at $T_m$  (see e.g.\ Ref.~\onlinecite{Sanna03}) and its observation confirms the homogeneity of the F doping. By contrast the additional upturn of $\lambda_T$ below 5\,K, visible in all samples, but larger in those without Fe order, has already been identified as an exclusive feature of the Sm ordering.\cite{Khasanov}

The sample $x$=0.085 deserves special attention: although ZF-$\mu$SR  shows that {\em all} implanted muons detect {\em static} magnetic order, it displays bulk superconductivity (see Fig.~\ref{fig:squid}). Incidentally susceptibility is unchanged down to 2K, i.e. superconductivity is not affected by the onset of magnetic order. This remarkable feature is incompatible with {\em macroscopic} clusters of SC and M phases, arising for instance from inhomogeneous F doping. Magnetic fields at a distance $d$ from an antiferromagnetic-type domain (with null macroscopic moment) vanish as $d^{-3}$ and a rough limit for the detection of a spontaneous internal field is 1 mT. Assuming a Fe magnetic moment of 0.3 $\mu_B$,\cite{Klauss} its dipolar field reaches this limit at $d \sim 1$ nm. Therefore SC domains much larger than 1 nm in radius would necessarily imply a distinct measurable muon fraction with vanishing internal fields, and a corresponding reduction of the magnetic fraction $f_M$. By contrast, a sizable FC susceptibility and $f_M\approx 1$ can arise from interspersed AF and SC clusters of nanoscopic size, reminiscent of what is observed in cuprates\cite{Sanna04,Savici,Russo,Kohsaka} and often referred to as {\em nanoscopic} coexistence. This dictates a superconducting mean domain size of a few nanometers, compatible with an average coherence length of order $\xi=2$ nm.\cite{Pallecchi} The nanoscopic coexistence and the suppression of the ordering temperatures, a typical finite-size effect, are both indicative of a phase separation with extremely small domains, which suggests a strong competition between the two order parameters.

Moving now on to $x$=0.10, best fits always require only relaxing fractions, with two decaying contributions $a_\mathrm{slow}$ and $a_\mathrm{fast}$. The first has a fairly constant relaxation rate $\lambda_\mathrm{slow}\!\!\approx\!\!0.2\,\mu s^{-1}$, while the second is temperature dependent. Its rate $\lambda_\mathrm{fast}$ (Fig.~\ref{fig:ZF}f) of course displays the upturn below $T=5$ K, related to Sm ordering.\cite{Khasanov} The absence of a simultaneous ordering of the Fe moments is however demonstrated by the modest value of this low temperature rate, that is much less than the Fe-related transverse rate $\lambda_T$ of the $x\le0.085$ sample, (see $x=0.085$ shown in the same plot). The $x$=0.15 and 0.20 samples exhibit the same moderate rates, corresponding to internal static fields not exceeding few tenths of mT, above 100 K, and few mT, just above 5 K. Therefore ZF-$\mu$SR consistently rules out any Fe magnetic ordering for $x\ge$0.10. Incidentally, the decomposition of the ZF-$\mu$SR signal into two components, $a_{slow}$ and $a_{fast}$,  persists up to high temperatures, and is due to static dipolar fields $B_d$ of the order of tenths of mT from the diluted ferromagnetic inclusions, already mentioned. Longitudinal field-dependent $\mu$SR confirms the static character of these small fields.

 \begin{figure}
\includegraphics[width=0.4\textwidth]{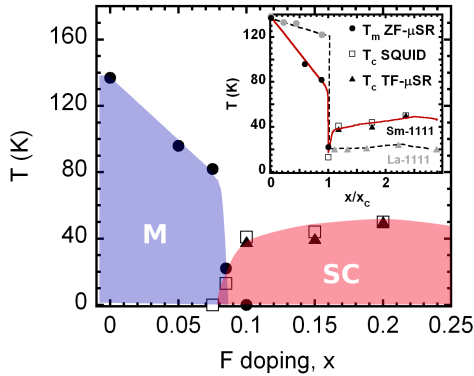}
 \caption{(Color online) Phase diagram of \sfaof; the areas are guides to the eye for magnetic (M) and superconducting (SC) regions. Inset: comparison of our data (solid line) to those of La-1111 from ref.[\onlinecite{Luetkens}] (dashed line), each rescaled to its nominal crossover concentration $x_c$.}
 \label{fig:PD}
 \end{figure}

Our results are summarized in the phase diagram of Fig.~\ref{fig:PD}. Electron doping by O-F substitution leads to a sharp drop of $T_m$ around a crossover concentration $x_c$=0.085, where the M-SC phase crossover occurs. Similar sharp boundaries in the phase diagram were recently observed by resistivity \cite{Hess} and nuclear resonant forward scattering \cite{Kamihara} The inset of Fig.~\ref{fig:PD} compares our data to those of La-1111,\cite{Luetkens} each rescaled to its nominal $x_c$ value: the behavior is strikingly similar. In both cases the magnetic order parameter suddenly collapses at $x_c$, and $T_c$ is only slightly dependent on doping. In the Sm sample the coexistence of magnetism with superconductivity is confined to an extremely narrow doping range, virtually point-like. The inset suggests that such a narrow coexistence region has probably escaped the La experiment, indicating that the two materials behave similarly at the M-SC crossover, both markedly differing from Co-doped 122 compounds.\cite{Laplace} Hence our results indicate that in 1111 materials the type of crossover does not depend on the rare-earth ion, in contrast with earlier reports,\cite{Drew,Zhao} which were most likely influenced by a much larger sample inhomogeneity.\cite{Amato,Uemura}

Concluding, our results unify the scenario of the crossover from magnetic order to superconductivity in R-1111 oxypnictides by offering a more consistent common picture. Thanks to the better chemical homogeneity of our samples we show that the behavior for R=Sm is compatible with that of the lower $T_c$ systems (R=La).  Both share a sharp crossover, evidence of a strong competition between the two order parameters. The fortunate availability of a homogeneous composition very close to the crossover ($x=$0.085), however, demonstrates a clear instance of nanoscopic coexistence, with reduced order parameters. This specific sample significantly indicates that also oxypnictides are prone to \textit{intrinsic} phase separation, a feature which is common to many unconventional superconductors.

This work was performed at the S$\mu$S, Paul Scherrer Institut, Villigen, Switzerland. We are grateful to the machine and beamline groups, whose outstanding efforts have made these experiments possible. We would like to thank Dr.~A.~Amato  and Dr. A.~Martinelli for support.

\end{document}